\documentclass[11pt]{article}

\usepackage{latexsym}
\usepackage{amssymb}
\usepackage{amsfonts}
\usepackage{amsmath}
\usepackage{bm}
\usepackage{centernot}
\usepackage{exscale,relsize}
\usepackage{multicol,ragged2e}  
\usepackage{blindtext}
\usepackage{stmaryrd}
\usepackage{dsserif}
\usepackage{hyperref}
\usepackage{cancel}
\usepackage{mathrsfs}
\usepackage[scr=rsfs,cal=boondox]{mathalfa}
\usepackage{multicol}
\usepackage[authoryear]{natbib}
\newtheorem{corollary}{Corollary}
\newtheorem{proposition}{Proposition}
\newtheorem{fact}{Fact}
\newcounter{exnum}%[section]

\newenvironment{definition}{\vspace{.1in}\noindent{\sc     	Definition:\/}}{{}}\vspace{.1in}

\newcounter{exnum1}
\newenvironment{remark}[1]{{\vspace{.1in} \noindent
{\sc Remark\addtocounter{exnum1}{1}
             \arabic{exnum1}}:
             }}
 {\vspace{.1in}}

\newcommand\bfzero{\mathbf{0}}
\newcommand\bfone{\mathbf{1}}

\newcommand\nvec[1]{\langle {#1} \rangle}

\newcommand\mfu{\mathfrak{u}}

\newcommand\mfz{\mathfrak{z}}
\newcommand\mfZ{\mathfrak{Z}}

\newcommand\be{\begin{equation}}
\newcommand\ee{\end{equation}}

\newcommand\ti{\underset{i}{\times}}

\newcommand\tI{\text{I}}

\newcommand\mbfwu{\mathbf{w}(u)}
\newcommand\G[1]{G_{\mbfwu}}
\begin{document}
  
\title{\bf Positively Homogeneous Saddle-Functions and Euler's Theorem  in Games}

\author{\large Joseph M. Ostroy\thanks{Department of Economics, UCLA: ostroy@ucla.edu}\,  and  Joon Song\thanks{Department of Economics, Sungkyunkwan University; joonsong.econ@gmail.com}}

\date{January 2025}
%\publishers{Very Big Publishers, New York}

%\uppertitleback{Upper Title Back}
%\lowertitleback{Lower Title Back}

%\dedication{Dedicated to my best friend.}

\maketitle \thispagestyle{empty}

\begin{abstract}
Connections are made between solution concepts for games in characteristic function form and Euler's Theorem underlying the neo-classical theory of  distribution in which the total output produced is imputed to the marginal products of the inputs producing it.  The assumptions for Euler's Theorem are constant returns (positive homogeneity) and differentiability of the production function. Representing characteristic functions in a vector space setting, marginal products of commodity inputs are translated as  marginal products of individuals.  Marginal products for discrete (resp.  infinitesimal) individuals are defined by discrete (resp. infinitesimal) directional derivates. Additivity of directional derivatives underlies the definition of differentiabilty in both discrete and infinitesmal settings. 

A key distinction is  between  characteristic functions defined by von Neumann and Morgenstern (vNM) which do not necessarily exhibit concavity and characteristic function that do. A modification of the definition, interpreted as introducing ``property rights,'' implies concavity. 

 The Shapley value is a {\em redefinition\/} of an individual's  marginal product for a (vNM) characteristic function. Concave characteristic functions  do not require such redefinition, and are closely related to the core. 
 
 Concave characteristic functions imply  the existence of positively homogeneous saddle-function functions whose saddle-points represent equilbria of the game.  The saddle-point property applies to games with populations consisting of any number of individuals each type. When there is a small integer number of each type, the saddle-point property is often, but not always, inconsistent with the Euler condition\,---\,that each individual receives its marginal contribution.   Conversely, the saddle-point condition is typically, but not always, consistent with the Euler condition when there are a large number of each type.  

The Euler condition   at a saddle-point  implies  that individuals have the incentive to report their utilities truthfully.     

\end{abstract}

%\footnotesize{t

%\noindent Economics:  Polyhedral quasi-linear and non-quasi-linear exchange. Walrasian equilibrium. 

%\noindent Games: Nash, Hannan, and Aumann (correlated) equilibrium. Identical interest and two-person zero-sum games. Market games. 

% \noindent Convex Analysis:  Subdifferential calculus, conjugate duality, bi-conjugacy, and  minimax,%}

\pagebreak

\thispagestyle{empty}

\tableofcontents

\thispagestyle{empty}

\pagebreak

\setcounter{page}{1}

\section{Introduction}

Departing from neo-classical economics, von Neumann and Morgenstern's model of cooperative games (\cite{vonNeumannMorgenstern1947}) makes no reference to commodities as the source of utility, while also adopting the hypothesis of transferable utility (TU). With TU, efficiency is defined as maximizing the sum of individual utilities. In the characteristic function formulation of a game, individuals can be regarded as the inputs producing utility as output. The focus of attention is on how total output is distributed among the individuals. Connections between cooperative games and the neo-classical marginal product theory of distribution used to explain the value of inputs is the focus, below. 

In the neo-classical formulation, ``marginal'' refers to the calculus of infinitesimals, while individual as inputs in a game are discrete. The infinitesimal margin is the limit of ratios of discrete differences\,---\,directional derivatives, allowing comparisons between the two margins. Transition from the discrete to the infinitesimal is achieved through replication of a discrete (finite) game, resulting in a constant return to scale production function. 
 
 The neo-classical marginal productivity theory of distribution stipulates that inputs are rewarded with the amounts they contribute to total output, their marginal products. An essential consistency condition, known as Euler's Theorem, is that the sum of the payments distributed should equal the total amount produced. In addition to the need for constant returns, Euler's Theorem also requires the production function to be differentiable -- formally, that directional derivatives are additive. Additivity of discrete differences, as well as additivity of directional derivatives in games, are highlighted, below.

 Section 2 includes a statement of Euler's Theorem and conditions for differentiability of positively homogeneous function. To provide a closer connection between Euler's Theorem and games, the characteristic function is restated as a real-valued function on a vector space, rather than on subsets of the set of individuals, allowing differences between sets to be represented as differences in vectors. In game theory terminology, a distinction is made between characteristic functions that are, or are not, equal to their superadditive covers. In vector terminology, the distinction is between characteristic functions that are, or are not, ``self-concavifying.'' Differences between the two classes of characteristic functions are emphasized, below.
 
  In Section 3, the vector formulation of a characteristic function allows the marginal products of individuals to be represented as discrete versions of a directional derivative. The Shapley value (\cite{Shapley1953}) for the not necessarily concave functions that underlie the characteristic functions defined by von Neumann and Morgenstern is restated in terms of the well-known revised definition of what constitutes an individual's marginal product. Exceptional properties for 2-person, and particularly constant sum games, are noted. Shapley's revised definition of marginal products for a finite game is extended to a positively homogeneous setting allowing for infinitesimal individuals and an Euler Theorem interpretation.
 
 Section 4 concerns the non-concavity/concavity of the characteristic function. Nonconcavity follows from the minimax definition of the gains a subset of individuals can achieve as the best it can do against the opposition of the complementary subset. A ``property rights'' modification of the characteristic function is given, eliminating the influence of the opposition. If a subset of individuals forms, it can access the activities/strategies it controls to maximize its joint gains without interference from those in the complementary subset. The modification is subject to the qualification that it does not apply to subsets consisting of pairs of individuals. The modification allows the self-concavity properties of 2-person games to be preserved without jeopardizing them when $n\geq 3$. The following sections are devoted to properties of the modified characteristic function.
 
 Section 5 defines positively homogeneous saddle-functions for games. A game in (modified) characteristic function form is viewed in the context of other games that vary with respect to the parameters defining them. The parameters are: the population of individuals of each of a finite number, $n$, of types and the utility functions defining each of the types, to be denoted respectively as a vector $x$ and a matrix $U$. Holding $U$ fixed, a function $F$ maximizing the gains with respect to $x$ is positively homogeneous and concave. Holding $x$ fixed, another function $H$ minimizing the gains is positively homogeneous and convex. A pair of solutions, one defining the valuations of individuals associated with the maximization of $F$ with respect to $x$ and the other minimizing solution for $H$ with respect to $U$, constitutes a saddle-point for the pair. The saddle-point exhibits Euler's Theorem when gradients exist for each function.
 
 Section 6 analyzes the properties of $F$ as it varies with $x$. Euler gaps are defined for individuals in games with discrete, i.e., finite numbers, and with infinitesimal individuals. The Euler condition is satisfied only when the gap is zero. Characterizations of zero and non-zero gaps are given, both for discrete and infinitesimal individuals. The exceptional cases of zero gaps with finite numbers and non-zero gaps for infinitesimal individuals are highlighted. Properties of characteristic functions defined by F are related to the existence of a non-empty core and to the Shapley value of such games, including conditions under which the saddle-points do, or not coincide with the core and the Shapley value. For 2-person games, a zero Euler gap exists if and only if the game is constant sum. 
 
 Properties of $H$ as it varies with $U$ are analyzed in Section 7. In the saddle-point construction, its properties mirror those in $F$. They are used to exhibit the incentive properties for games with zero and non-zero Euler gaps. Games with zero (resp. non-zero) Euler gaps exhibit the presence (resp. absence) of incentive compatibility.
 
 In addition to the above-mentioned departure from neo-classical economics that games make no reference to commodities as a source of utility, in games individual utilities also depend on the actions of others rather being limited to the commodities individuals directly consume as ``private goods.'' When the private goods assumption is adapted to define a game in characteristic function form, the property rights feature of the characteristic function is well-known to be satisfied. Indeed, the property rights model above is an extension of the private goods assumption to accommodate externalities within, but not between, groups of individuals. Section 8 defines the characteristic function for a private goods exchange economy having the same finite number of types and the same finite number of activities as a standard game in normal form. Such private goods games exhibit saddle-functions with properties identical to those established in Sections 5 through 7.

 Analysis is confined to the original setting for games with a finite number (of types) of  individuals, each of which has a finite set of actions/strategies. In the concluding Section 9, brief comments are made about existing and conjectured extensions of the above results.

\section{Preliminaries}

\subsection{Euler's Theorem in Economics}

 {\em Euler's Theorem} characterizes the properties  of a positively homogeneous function, known in economics as {\em constant returns to scale}.  A production function $f$ exhibiting constant returns  is regarded as freely replicable and, therefore, all of the output is attributable to the inputs producing it. This formulation underlies the marginal productivity theory of distribution in economics. Each infinitesimal unit of an input receives the extra gains it contributes to the total as if it were the last unit employed. 
 
With constant returns, 
\[f(\lambda x) = \lambda f(x), \lambda >0\] 
 Assuming $f$  is  differentiable\,---\,a standard assumption in neo-classical economics, payments to homogeneous, divisible factors of production, $x=(x_1,\ldots,x_n)$, based on the vector of marginal products defined the gradient $\nabla f(x)$ add up to the output they produce,  
 \[  \nabla f(x)\cdot x = f(x). \]

A reframing of Euler's Theorem translates the differential calculus of the marginal products of commodity inputs in neo-classical economics to the marginal products of individuals in cooperative games.  The presence of non-differentiability in activity analysis models of constant returns with respect to commodity inputs is related to non-differentiability with respect to individuals associated with polyhedral properties of games. 

\subsection{Differentiability of a Positively Homogeneous Function}

For $f:\mathbb{R}^n_+ \rightarrow \mathbb{R}_+$ and $x \in \text{int}\, \mathbb{R}^n_+$, the directional derivative at $x$ in the direction $y\in \mathbb{R}^n$ is
\[ f(x)(y) := \lim_{\lambda \searrow 0} \frac{f(x +\lambda y) - f(x)}{\lambda} \]
The gradient of $f$ at $x$, $\nabla f(x)$,  exists if and only if 
\[ -f(x)(-x_i) = f(x)(x_i), \quad i =1,\ldots, n\]

%$g$ is positively homogeneous when
%\[ g(\lambda x) = \lambda g(x), \quad x \in \mathbb{R}^n_+, \lambda > 0\]

With positive homogeneity, there is a ``one-sided'' test of differentiablity.

\begin{proposition} If $f$ is positively homogeneous, 
\[ \sum_i -f(x)(-x_i) = f(x) \Longrightarrow -f(x)(-x_i) = f(x)(x_i),  i =1,\ldots, n \Longrightarrow \nabla f(x)\cdot x = f(x)\]
\end{proposition}

\subsection{Characteristic Functions on $\mathbb{R}^n_+$} 

The characteristic function of a game $\phi:\mathcal{S} \rightarrow \mathbb{R}_+$, where $\mathcal{S} = \{S: S\subseteq \tI\}$ is the set of subsets of $\tI= \{1,\ldots,n\}$ including $S = \emptyset$ and $\phi(\emptyset) = 0$. The function $\phi$ includes the superadditivity condition
\[ S\cap T = \emptyset \Longrightarrow \phi(\bfone_{S\cup T}) \geq \phi(\bfone_{S}) + \phi(\bfone_T)\]

The set of subsets $\mathcal{S}$ is represented by vectors in $\mathbb{R}^n_+$. $\bfone_{\tI} = (1,1,\ldots,1) \in \mathbb{R}^n$, and $\bfone_S(i) = 1$ if $i\in S$, $=0, i\notin S$;  $S= \emptyset, \bfone_S =\bfzero$.  The domain of the characteristic function is
\be \bfone_{\mathcal{S}} = \{\bfone_{S}: S\subseteq \tI\}\ee
 The set function $\phi: \mathcal{S} \rightarrow \mathbb{R}_+$ is represented as the function  $ G: \bfone_{\mathcal S} \rightarrow \mathbb{R}_+$,
\be G(\bfone_S) = \phi(S), \forall S \ee
An alternative set function $\Phi:\mathcal{S}\rightarrow \mathbb{R}_+$, represented by  $F: \bfone_{\mathcal S} \rightarrow \mathbb{R}_+$ is 
\be \Phi(S) = F(\bfone_S) = \sup\, \{\sum_k \alpha_k G(\bfone_{S_k}) : \alpha_k \geq 0, \sum_k \alpha_k \bfone_{S_k} = \bfone_S\} \geq G(\bfone_S) = \phi(\bfone_S)\ee

In game theory terminology, $F$ is called the superadditve cover of $G$. Regarding $G$ and $F$ as functions on $\mathbb{R}^n_+$,  $F$ is the smallest concave (polyhedral) function $\geq G$ on $\bfone_{\mathcal{S}}$.  
 When $G= F$ on  $\bfone_{\mathcal S}$, $G$ can be said to be {\em self-concavifying}.

Comparisons between $\phi \sim G$ and $\Phi \sim F$ are the focus of attention, below.

\section{Euler's Theorem and the Shapley Value for $G$}

\subsection{Discrete Version}

As a discrete directional derivative of $G$ at $\bfone_{\tI}$, the contribution of $i$ to $\tI$ is
\[-\Delta G(\bfone_{\tI})(-\bfone_i) := - [ G(\bfone_{\tI} -\bfone_i) -  G(\bfone_{\tI}] = G(\bfone_{\tI}) - G(\bfone_{\tI} - \bfone_i)\]

\begin{definition} The {\em discrete Euler Condition for $G$ at\/} $\bfone_{\tI}$ is that the sum of the individual contributions equals the total gains:
\[\sum_{i\in I} -\Delta G(\bfone_{\tI})(-\bfone_i)] = G(\bfone_{\tI}) \]
\end{definition}

\begin{proposition} When $n=2$,  $G =F$.  And either $G$ is strictly superaddititive 
\[ -\Delta G(\bfone_{\tI})(-\bfone_1) + -\Delta G(\bfone_{\tI})(-\bfone_2)  > G(\bfone_{\tI}) >  G(\bfone_2) + G(\bfone_1)  \]
or $G$ is additive, i.e., constant sum, in which case the discrete Euler conditon holds:
\[ -\Delta G(\bfone_{\tI})(-\bfone_1) + -\Delta G(\bfone_{\tI})(-\bfone_2)  =G(\bfone_{\tI}) =  G(\bfone_2) + G(\bfone_1)\]
\end{proposition}

With that important exception, when $n\geq 3$,  $G \leq F$.    a characteristic function may satisfy
 \[ -\Delta G(\bfone_{\tI})(-\bfone_1) + -\Delta G(\bfone_{\tI})(-\bfone_2) +-\Delta G(\bfone_{\tI})(-\bfone_3) <  G(\bfone_{\tI}) \]

\subsection{Shapley's Transformation of $G$}

\begin{definition} The contribution of $i$ to $\tI$  is {\em re-defined\/} as a discrete directional derivative of $G^{\text{Sh}}$ at $\bfone_{\tI}$ based on $G$, as
\[-\Delta G^{\text{Sh}}(\bfone_{\tI})(-\bfone_i)  := \sum_{S\in \mathcal{S_i}} \alpha_S [G(\bfone_{S\cup \{i\}}) - G(\bfone_{S})]\] 
where $\mathcal{S}_i = \{S: i\notin S\}$ and
 \be\alpha_S = \frac{|S|!(|\tI|-|S-1|)!}{|\tI|!} \ee
\end{definition}

From this, Shapley showed 
\be \sum_{i\in I} -\Delta G^{\text{Sh}}(\bfone_{\tI})(-\bfone_{i}) = G(\bfone_{\tI})\ee

\begin{remark} {} An added property of $G^{\text{Sh}}$ is that it satisfies the discrete Euler condition for each  $G^{\text{Sh}}(\bfone_{S})$. Define 
\be   -\Delta G^{\text{Sh}}(\bfone_S)(-\bfone_{i}) = \sum_{T\in \mathcal{T}_i(S)} \alpha_T [G(\bfone_T\cup\{i\}) - G(\bfone_T)]\ee
where $\mathcal{T}_i(S) = \{T\subset S: i\notin S\}$,
 \be \alpha_T = \frac{|T|!(|S|-|S-1|)!}{|S|!}\ee 
The Euler version of the Shapley value for the game defined by $G(\bfone_T), \forall T\subseteq S$ is
\be \sum_{i\in S} -\Delta G^{\text{Sh}}(\bfone_S)(-\bfone_{i}) = G^{\text{Sh}}(\bfone_S)\ee
\end{remark}

\subsection{Gradient Version of the Shapley Transformation} 

Let $G^{\text{Sh}}_{\infty}: \mathbb{R}_+^n \rightarrow \mathbb{R}_+$ be a positively homogeneous extension of $G^{\text{Sh}}$: 
\[   G^{\text{Sh}}_{\infty}(\lambda x) =  \lambda G^{\text{Sh}}_{\infty}(x), \quad x\in \mathbb{R}^n_+, \lambda > 0\]
with
\[G^{\text{Sh}}_{\infty}(\bfone_S) = G^{\text{Sh}}(\bfone_S) = G(\bfone_S), \quad \forall S\]
 
\begin{definition} The directional derivative of $G^{\text{Sh}}_{\infty}$ at $\bfone_{\tI}$ in the direction $-\bfone_i$ is
\[ G^{\text{Sh}}_{\infty}(\bfone_{\tI})(-\bfone_i) := \lim_{\lambda \searrow 0} \frac{G^{\text{Sh}}_{\infty}(\bfone_{\ti} - \lambda \bfone_i) - G^{\text{Sh}}_{\infty}(\bfone_{\tI})}{\lambda}\]
\end{definition}

From positive homogeneity, 
\[ -  G^{\text{Sh}}_{\infty}(\bfone_{\tI})(-\bfone_{\tI}) =  G^{\text{Sh}}_{\infty}(\bfone_{\ti})(\bfone_{\tI}) = G^{\text{Sh}}_{\infty}(\bfone_{\ti})\]
Therefore,
\be -  G^{\text{Sh}}_{\infty}(\bfone_S)(-\bfone_S) =  G^{\text{Sh}}_{\infty}(\bfone_S)(\bfone_S) = G^{\text{Sh}}_{\infty}(\bfone_S) = G(\bfone_S), \quad \forall S\ee

Ordinarily, the value of the ratio
\be  \frac{G^{\text{Sh}}_{\infty}(\bfone_{\tI}) - G^{\text{Sh}}_{\infty}(\bfone_{\tI}- \lambda \bfone_i)}{\lambda}  
= G^{\text{Sh}}_{\infty}( \lambda^{-1} \bfone_{\tI}) - G_{\infty}^{\text{Sh}}(\lambda^{-1} \bfone_{\tI} -  \bfone_i) \ee
in the definition of the (infinitesimal) directional derivative of a positively homogeneous function varies with $\lambda >0$. Here, (\ref{infinitesimal directional derivative})  is defined by  
  the discrete directional derivative
\be \label{infinitesimal directional derivative}-G_{\infty}^{\text{Sh}}(\bfone_{\tI})(-\bfone_i)  := \lim_{\lambda \searrow 0} G^{\text{Sh}}_{\infty}( \lambda^{-1} \bfone_{\tI}) - G_{\infty}^{\text{Sh}}(\lambda^{-1} \bfone_{\tI} -  \bfone_i) 
 =  - \Delta G^{\text{Sh}}(\bfone_{\tI})(-\bfone_i). \quad 0 < \lambda \leq 1\ee

Additivity of directional derivatives defines a literal version of Euler's Theorem
$\nabla G^{\text{Sh}}_{\infty}(\bfone_{\tI}) = \nvec{-G^{\text{Sh}}_{\infty}(\bfone_{\tI})(-\bfone_i)}$
\[  \nabla G^{\text{Sh}}_{\infty}(\bfone_{\tI}) \cdot \bfone_{\tI} = G^{\text{Sh}}_{\infty}(\bfone_{\tI}) = G(\bfone_{\tI})\]

The positively homogeneous extension  $G_{\infty}^{\text{Sh}}$ of $G$ will be compared  to the positively homogeneous extension of $F$, where the ratio comparable to (\ref{infinitesimal directional derivative}) varies with $\lambda$.

\begin{remark} {} The interpretation of the Shapley Value as Euler's Theorem is consistent with a special case of results in (\cite{AumannShapley1974}) where the number of types is finite.
\end{remark}

\section{Normal Form  of the  Characteristic Function}

This section  contrasts the properties  of $\phi (S) = G(\bfone_S)$  based on the standard derivation of  the characteristic function from the normal form of a game with  an alternative ``property-rights'' version  of the normal form where $\Phi (S) = F(\bfone_S)$ and $\Phi(\tI) = \phi (\tI)$.  

\subsection{Origin of $\phi\sim G$}

The set of pure strategies in an $n$-person game is
\[A = A_{\tI} = A_1\times A_2\times \cdots \times A_n\]
The set available to $S$ is $A_S = \times_{i\in S} A_i$

Convex combinations of $A_S$ are 
\[ \mfZ_S = \{\mfz_S : \mfz(a_S)\geq 0, \sum_{a_S \in A_{S}} \mfz(a_S) =1\} \quad S\subseteq {\tI} \]
The tensor product of $\mfz_S$ and $\mfz_{-S}$ is $\mfz_{-S} \otimes \mfz_{-S}$. The indicator function of $\mfZ_S\otimes \mfZ_{-S}$ is 
\[ \delta_{\mfZ_S\otimes \mfZ_{-S}}( \mfz_S \otimes \mfz_{-S}) = \begin{cases}  0 & \text{ $\mfz_S\in  \mfZ_S, \mfz_{-S} \in \mfZ_{-S}$}\\
                  = \infty, \text{otherwise}
\end{cases}                                                                                         
\]

When $S= \tI$, 
\be \mfZ_{\tI}  = \mfZ = \{\mfz  : \mfz(a) \geq 0,  \sum_a \mfz(a) =1 \}  \ee
%[This is to highlight connection with (?) discrete differentiability, below.]

%...First derive $\mathbf{v}_{\mfu}(S)$  as the total gains to $S$....then use LE as a decentralized pricing method of distributing the gains. This gives added credibility to this method of distribution as the implication of differentiabilty. 

Individual $i$'s utility is represented by the vector $\mfu_i \in \mathbb{R}_+^A$ (normalized to be non-negative).  The matrix with $n$ rows and $|A| = |A_1| + |A_2| + \cdots + |A_n|$ columns
\[ U = \nvec{u_i} \in \mathbb{R}_+^n\times  R^A_+\]
represents the utilities of the $n$ individuals in a game. 

The sum of utilities for $S$ is represented in vector-matrix notation as
\[ \bfone_S U := \sum_{i\in S} \mfu_i \in \mathbb{R}^A_+\]
The standard definition of the value of the characteristic function for $S$ is
\be \phi (U)(S) = \inf_{\mfz'_{-S}} \sup_{\mfz_S} \sum_{\mfz_S}  \, \{\bfone_S U\mfz_{-S})\}(\mfu_S) - \delta_{\mfZ_s\otimes \mfZ_{-S}}( \mfz_S \otimes \mfz_{-S}) \ee

$\phi(U)(\cdot )$ is superadditive on $\mathcal{S}$:  
\be \phi(U)(S\cup T) \geq \phi(U)(S ) + \phi(U)(T) \quad S\cap T = \emptyset \ee
For  ${\tI}$,
\be \phi (U) (\tI) := \delta_{\mfZ}^*(\bfone_{\tI}U) = \sup_{\mfz}  \,  \{\bfone_{\tI}U  \cdot \mfz - \delta_{\mfZ_{\tI}}(\mfz)\}  \ee

\subsection{From $\phi \sim G$ to $\Phi \sim F$: Self-concavifying Characteristic Function}

The definition of the characteristic function includes the possiblity that 
\be \label{non-natural characteristic function}\sup\, \big\{\sum_k \alpha_{S_k} \phi(U)(S_k): \alpha_{S_k} \geq 0.  \sum_k \alpha_{S_k} \bfone_{S_k} = \bfone_S\big\} > \phi(U)(S)\ee

To preclude (\ref{non-natural characteristic function}), the set of utility functions determining $\phi(U)$ could be restricted to those
$\overset{\circ}{U} = (\overset{\circ}{u}_1,\overset{\circ}{u}_2, \ldots, \overset{\circ}{u}_n)$ such that
\be  \label{natural characteristic function}\sup\, \{\sum_k \alpha_{S_k} \phi(\overset{\circ}{U})(S_k): \sum_k \alpha_{S_k}\bfone_{S_k} = \bfone_S\} = \phi(\overset{\circ}{U})(S), \forall S \in \mathcal{S} \ee
%.......(later) Note: For two-person  games, this is not a binding constraint, i.e., $\overset{\circ}{U} = U$.

To ensure that (\ref{natural characteristic function}) holds for all $U$, the definition of the characteristic function is changed to include a {\em collective\/}, rather than an individual, property rights restriction.   

For $\delta_{\mfZ_S}$, the characteristic function of $\mfZ_S$ when $|S|\geq 3$, define
\be \Phi(U)(S) = \delta^*_{\mfZ_S}(\bfone_S U) = \sup_{\mfz}\, \{\bfone_S U\cdot \mfz - \delta_{\mfZ_S} (\mfz)\}\ee
It differs from 
 \[\phi(U)(S) = \inf_{\mfz_{-S}} \sup_{\mfz_S}\, \{\bfone_S U\cdot \mfz_S(\mfz_{-S}) - \delta_{\mfZ_S(\mfz_{-S})}(\mfz_S)\}\] 
exactly because the value $\bfone_S U\cdot \mfz$ is restricted to $\mfz\in \mfZ_S$ rather than $\bfone_S U\cdot \mfz_S\otimes \mfz_{-S}$. I.e., it gives   $S$ {\em exclusive access\/} to $A^S$, thereby precluding $\tI\backslash{S}$ from interfering. 

{\em For $|S| \leq 2$, the standard definition applies.} The qualification is adopted to preserve the properties for all $2$-person games and $2$-person subgames when $n\geq 3$ where $G(\bfone_S') =F(\bfone_S')$, $S' \in \{ i {ij}\}$, while not allowing them to interfere when $S \neq S'$.

The revised definition gives a self-concavified characteristic function implying there are no gains to convex subdivisions of $S$.   
\be
\begin{aligned}
 \Phi(U)(S)  = F(\bfone_S) = \delta^*_{\mfZ_S}(\bfone_SU) = \sup\, \{\sum_k \alpha_{S_k} \delta^*_{\mfZ_{S_k}}(\bfone_{S_k} U) : \alpha_{S_k} \geq 0,  \sum_k \alpha_{S_k} \bfone_{S_k}
  = \bfone_S\}\\
  \end{aligned}
\ee
The modified characteristic function has the same gains to $\tI$ as the original, 
\be \phi(U)(\tI) = \Phi(U)(\tI) = F(\bfone_\tI)\ee

%\be 
%\begin{aligned}
%F_U(x) & = \sup\, \big\{ \sum_k \alpha_{S_k} \delta^*_{\mfZ_{S_k}}(\bfone_{S_k}U) :  \alpha_{S_k} \geq 0, \sum_k \alpha_{S_k}\bfone_{S_k}  = x\big\} \\
 %& =\sup\, \big\{ \sum_k \alpha_{S_k} F_{U}(\bfone_{S_k}) :  \alpha_{S_k} \geq 0, \sum_k \alpha_{S_k}\bfone_{S_k}  = x\big\}
%\end{aligned}
%\ee
As an objective function, $F(x)$ is an allocation problem in which the set of possible ``communities,'' $S_k\subseteq \tI$, each requires the participation of $\bfone_{S_k}$ that is limited to the ``resource constraints'' $\mfZ_{S_k}$, achieves the greatest possible total gains when the total population is $x$. 

\begin{remark} {}  {\sc (Public Goods)} When $F(\bfone_{\tI})$ is regarded as a model with one individual of each type,  interdependent utilities $\mfu_i(a_1,a_2, \ldots, a_n)$ can be interpreted as a model of (pure) public goods to exploit the gains from $\bfone_{\tI}U = \sum_{i\in \tI} u_i$. 
However, with 
$F(k\bfone_{\tI})$, where multiple communities $S_k\subset \tI$  may form, it becomes a model of local public goods with {\em collective property rights\/} for each $S_k$ with the possibility that $i$ might join any $S_k$ containing $i$. Equivalently, it is a ``group matching problem'' allowing members of each $S_k$ to choose among $\mfZ_{S_k}$ to convexify their possible choices. 
\end{remark} 

\begin{remark} {}  {\sc (Private Goods)/Externalities}  Precedent for the self-concavifying modification of the characteristic function is found in the adaptation of games to accommodate the private goods models of exchange, where the property rights assumption is well-known to fit  without qualification. See Section 9.
\end{remark} 

%The property rights assumption in exchange is guaranteed by the hypothesis that (i) each $i$ has a given allocation of commodities $\omega_i \in \mathbb{R}^{\ell}_+$ and  the utilities of  $S$ are limited to those that can be achieved by allocations $y_i \in  \mathbb{R}^{\ell}_+$ such that $\sum_{i\in S} y_i = \sum_{i \in S} \omega_i$; hence, by constraints that are independent of $\tI\backslash{S}$. In addition, (ii) the utility of $i$ is $\nu_i(y_i)$, depending only on the individual allocation of commodities to each $i$.  The public goods version dispenses with (ii).

\section{The Saddle-Function Underlying $\Phi(U)\sim F_U$}

To call attention to the dependence of total gains on utilties, write $F(x)$ as 
\[  F_U(\bfone_S) = \sup_{\mfz}\, \{\bfone_S U\cdot \mfz - \delta_{\mfZ_S}(\mfz)\}\]

For $x \in \mathbb{R}^n_+$,  let $ \text{supp}\, x := \{i\in \tI: x_i >0\}$ and
\[\mfZ_x = \{\mfz \in \mfZ_{\tI}: \sum_{a^S} \mfz (a^S) =0, S \not\subseteq  \text{supp}\, x \}\]

Extend $ F_U(\bfone_S)$ to 
\be F_U(x) := \sup_{\mfz}\, \{ xU\cdot \mfz - \delta_{\mfZ_x}(\mfz)\} \ee  

In addition, let
\be 
H_x(U) := \inf_{\mfz}\, \{ xU\cdot \mfz - \delta_{\mfZ_x}(\mfz)\} 
\ee

%Adopting the self-concavifying characteristic function $F_U$ yields the following duality properties. 

\begin{proposition} Holding $U$ fixed,  $F_U(x)$ is positively homogeneous  
and subadditve, 
\[ F_U(\lambda x) = \lambda F_U(x), \lambda >0 \qquad F_U(x + x') \leq F_U(x) + F_U(x')\]
therefore concave on  $\mathbb{R}^n_+$. 
\end{proposition}

Positive homogeneity is interpreted as a constant return to population. Let $\mfz(x):=\arg\sup \{xU\cdot \mfz - \delta_{\mfZ_x}(\mfz)\}$. Subadditivity derives from the fact that if $\mfz(x)\neq \mfz(x')$ is allowed, the value can be maximized more efficiently, compared to the case that $\mfz(x)=\mfz(x')$ is required.  

\begin{proposition} Holding $x$ fixed,  $H_x(U)$ is positively homogeneous and superadditive, 
  \[ H_x(\lambda U) = \lambda H_x(U), \lambda > 0 \qquad H_x(U +U') \geq H_x(U) + H_x(U')\]
therefore convex on  $(\mathbb{R}^n_+ \times \mathbb{R}^A_{+})$.  
\end{proposition}

Similarly, scaling individual utilities does not change argmin of $H_x(U)$. Superadditivity derives from the fact that the minimizers for the right hand side do not have to be identical. 

%...subadditive:  value from argmin for $U+U'$ combination of utilities for $x$ not smaller than  values from separate argmins for $U$ at $x$ and $U'$ at $x$. 
 
\begin{definition} The {\em subdifferential of the polyhedral concave function} $F_U$ at $x\in \mathbb{R}^n_+$ is
\[\partial F_U(x) := \{ r : r\cdot x - F_U(x) \geq  r\cdot x' - F_U(x'), \forall x' \in \mathbb{R}_+^n\}\]
Because $F_U$ is  positively homogeneous, $r\in \partial F_U(x)$ implies
\be 0=  r\cdot x - F_U(x) \geq  r\cdot x' - F_U(x'), \forall x' \in \mathbb{R}_+^n\ee
\end{definition}

\begin{fact} 
  $\partial F_U(x)$ is a non-empty, polyhedral convex set 
  \end{fact}

\begin{definition} The {\em subdifferential of the polyhedral convex function} $H_x$ at $U$ is 
\[ \partial H_x(U) = \{\mfz:   xU\cdot \mfz - H_x(U) \leq  xU'\cdot \mfz  - H_x(U'), \forall U' \in (\mathbb{R}^A_+\times \mathbb{R}^A_+)\] 
Because $H$ is positively homogeneous, $\mfz \in \partial H(U)$ implies
\be  xU'\cdot \mfz  - H_x(U') \geq   xU\cdot \mfz - H_x(U) = 0 \quad \forall U' \in (\mathbb{R}^A_+\times \mathbb{R}^A_+) \ee
\end{definition}

In parallel with $\partial F_U(x)$, 
\begin{fact} $\partial H_x(U)$ is a non-empty, polyhedral convex set.
\end{fact}

For $U\in \mathbb{R}_+^n \times \mathbb{R}^A_+$ and $\mfz \in  \mathbb{R}^A_+$, $U\mfz \in \mathbb{R}^n_+$ and $x\cdot U\mfz  = xU\cdot \mfz \in \mathbb{R}_+$. The transpose of  $U\mfz$ is $\mfz^t U^t \in \mathbb{R}^n_+$  and
\be xU\cdot \mfz =\sum_a [\sum_i x_i  u_i(a)] \cdot \mfz(a) = \sum_i [\sum_a \mfz(a) u_i(a)] \cdot x_i 
 = \mfz^t U^t \cdot x\ee

Existence of subdifferentials defines a saddle-point for the positively homogeneous saddle-function  defined by the parameters $(x,U)$: Let $z(x,U) \in \partial H_x(U)$ and $r(x,U) \in \partial F_U(x)$, $r(x,U) = \mfz(x,U)^tU^t$. Then 
\be 
\begin{aligned}
& x U'\cdot \mfz(x,U) - H_x(U')  \geq  xU\cdot \mfz(x,U) - H_x(U) \\
&  = xU\cdot \mfz(x,U) - F_U(x)  = 0 = \mfz(x,U)^t U^t\cdot x -F_U(x)\\  
&  = r(x,U)\cdot x - F_U(x)  \geq r(x,U) \cdot x' - F_U(x')  
   \end{aligned}
\ee

The set of saddle-points for $(x,U)$ is
\be  \big(r(x,U),\mfz(x,U)\big)\in \partial F_U(x)\times \partial H_x(U) \ee
I.e., {\em any $\mfz(x,U) \in \partial H_x(U)$ can be paired with any} $r(x,U) = \mfz(x,U)^t U^t \in F_U(x)$ to form a saddle-point. 
Compatibility of the pairing is achieved by ``money transfers'': there is a one-to-one correspondence between $\mfz(x,U) \in \partial F_U(x)$ and $\mfz(x,U)^t U^t \in \partial H_x(U)$ defined by
\be m_i(\mfz(x,U)) = u_i\cdot \mfz(x,U) - r_i(\mfz(x,U)) \ee
By construction
\be m(\mfz(x,U)) \cdot x = \sum_i [u_i\cdot \mfz(x,U) - r_i(\mfz(x,U))]x_i = xU\cdot \mfz(x,U) - \mfz(x,U)^t U^t\cdot x =0 \ee
The money transfer $m_i(\mfz(x,U)) \gtreqless 0$ converts $u_i\cdot \mfz(x,U)$\,---\,the gains  $i$ receives {\em directly} from  $\mfz(x,U)$\,---\,to $i$'s net gains,  $r_i(x,U)$.

The ``flexibility'' of money transfer allows
\begin{proposition} (\sc{Subdifferential Characterization of  Positively Homogeneous Saddle-points})
\[ \partial F_U(x)\cdot x = F_U(x) = H_x(U) = \partial H_x(U) \cdot U\]
\end{proposition}

Differentiability of concave and convex functions are characterized by subdifferentials that are singletons. 
\begin{fact} $\nabla F_U(x) \neq \emptyset \Longleftrightarrow \{r\} = \partial F_U(x)  \Longleftrightarrow   \{\mfz\} = \partial H_x(U) \cdot U   \Longleftrightarrow \nabla H_x(U) \neq \emptyset $
\end{fact}

 With differentiabililty money transfers are unquely determined by  $\nabla H_x(U)$ or $\nabla  F_U(x)$.  Absence (resp. presence) of indeterminacy in money transfers characterizes existence (resp. non-existence) of Euler's Theorem for positively homogeneous saddle-functions.  
\begin{proposition}  (\sc{Euler's Theorem of the Positively Homogeneous Saddle-point})
\[ \nabla  F_U(x)\cdot x = F_U(x) \Longleftrightarrow \nabla H_x(U)\cdot U = H_x(U)\]
\end{proposition}

Additional relevant properties of the subdifferentials of $F$ and $H$ are exhibited, below.
Application of saddle-point existence for $n=2$ and the special case that $U$ is constant sum are treated separately in the following Section. 
(See \cite{Rockafellar1970}, ch 7, for more general formulations of saddle functions.)

\section{Properties of $F_U$} 

 \subsection{Euler Gaps}
 
 Since $U$ is fixed throughout this section, $F_U$ is denoted by $F$.

Indivisibility of individuals is not a binding constraint for $F$.  For any integer valued $x$. $F(x)$ is both 
the maximum gains when there are an integral number of individuals of each type; and it is also the gains when each $x_i$ is regarded as mass of infinitesimal individuals of type $i$. Therefore, the same setting can be used to define  the ``calculus'' for discrete and infinitesimal individuals. 

The discrete directional derivative of the positively homogeneous function $F$ at $x$ in the direction $d$ is
\be \Delta_k F(x)(d) := \frac{F(x+k^{-1}d) - F(x)}{k^{-1}} = F(kx +d) -F(kx), \quad k = 1,2,\ldots\ee
with $k=1$ the most discrete case.  With positive homogeneity, the infinitesimal version is the limiting value of the discrete version $\Delta_k F(x)(d)$ as the scale increases.  
\be F(x)(d) = \lim_{k\rightarrow \infty} \Delta_k F(x)(d)\ee 

The focus is on $x = \bfone_{\tI}$ and $d = \bfone_i$ and 
\be  - \Delta_k F(\bfone_{\tI})(-\bfone_i) = \frac{-[F(\bfone_{\tI} - k^{-1}\bfone_i) - F(\bfone_{\tI})]}{k^{-1}} =  F(k\bfone_{\tI}) -F(k\bfone_{\tI} - \bfone_i) \ee
The properties of $F$ imply that the sum of the individual contributions added separately are at least as large as their joint contribution.
\be \sum_i [- \Delta_k F(\bfone_{\tI})(-\bfone_i)] \geq - \Delta_k F(\bfone_{\tI})(-\bfone_{\tI}) = F(k\bfone_{\tI}) -F(k\bfone_{\tI} - \bfone_{\tI}) = F(\bfone_{\tI}) \ee

\begin{definition} The {\em discrete Euler Gap\/} for $F$ at $k\bfone_{\tI}$ is
\[ \mathcal{E}_k (\bfone_{\tI}) =  \sum_i [- \Delta_k F(\bfone_{\tI})(-\bfone_i)] - F(\bfone_{\tI}) \geq 0\]
The {\em infinitesimal  Euler Gap\/} for $F$ at $\bfone_{\tI}$ is
\[ \mathcal{E}_{\infty} (\bfone_{\tI}) = \sum_i [- F(\bfone_{\tI})(-\bfone_i)] - F(\bfone_{\tI}) \geq 0\]
\end{definition}
The term ``gap'' applies when the value is strictly positive. When it is zero, there is no gap.
 
\begin{proposition} ({\sc Discrete Characterization of the Euler Condition})\\
The following are equivalent:
\begin{enumerate}
 \item [(i)] $\mathcal{E}_k (\bfone_{\tI}) =0$
 \item [(ii)] $\nabla F(\bfone_{\tI}) =\{r\} = \partial F(k\bfone_{\tI})$
\end{enumerate}
\end{proposition} 

\begin{proposition} ({\sc Infinitesimal  Characterization of the Euler Condition})
\begin{enumerate}
\item [(i)] $\mathcal{E}_{\infty}(\bfone_{\tI}) = 0$
\item [(ii)] $\nabla F(\bfone_{\tI}) =\{r\}  = \partial F(\bfone_{\tI})$
\end{enumerate}
\end{proposition}

From (ii) in Propositions 7 and 8,
\begin{corollary} The discrete Euler condition implies the infinitesimal.
\end{corollary}

The polyhedral property of $F$ implies a qualified converse.

\begin{proposition} {(\sc Polyhedral Differentiability and non-Differentiability)}\\
(i) $\nabla F(\bfone)\neq \emptyset$   only if  there exists a positive integer  $K$ such that for all $k = K, K+1,K+2, \ldots $,
\[ \mathcal{E}_k(\bfone_{\tI}) = 0 \]
(ii) $\nabla F(\bfone) = \emptyset$ only if there exists an $\epsilon >0$ such that 
\[\mathcal{E}_k(\bfone_{\tI}) > \epsilon, \forall k\]
\end{proposition}

Differentiability requires that $F$ is infinitesimally ``flat'' at $\bfone_{\tI}$. However, because $F$ is  polyhedral, infinitesimal flatness implies  non-infinitesimal flatness.

For (ii),  $F$ is ``kinked''  at $\bfone_{\tI}$, i,e., $\partial F(\bfone_{\tI})$ is not a singleton, $E_{\infty}(r) >0$. Then  {\em all\/} finite approximations are bounded away from $0$.

\begin{remark} {} If $F$ were concave but not polyhedral, the equality in Proposition 6 need not be satisfied for any $K$. Nevertheless, the definition of the directional derivative says that $[F(\bfone_{\tI}+ \lambda d) - F(\bfone_{\tI})]$ converges to $F(\bfone_{\tI})$ 
{\em faster than\/} $\bfone_{\tI}+ \lambda d$ converges to $\bfone_{\tI}$. Therefore, if $\nabla F(\bfone_{\tI}) = \nvec{ \nabla F(\bfone_{\tI}; \bfone_i)}$ exists,   approxmations involving smaller and smaller indivisible units become  good approximations to the limiting case of infinitesimal units. To illustrate, let $\lambda_k = k^{-1}$ and $\nabla F(\bfone_{\tI}) = \nvec{\nabla F(\bfone_{\tI};\bfone_i)}$.  For every $\epsilon >0$, there exists $K$ such that if $k >K$,
\be   \frac{|\nvec{F(\bfone_{\tI}) - F(\bfone_{\tI} - k^{-1}\bfone_i )}- \nabla F(\bfone_{\tI};\bfone_i)|}{k^{-1}} < \epsilon \label{k approximation}\ee
If  $F$ is not differentiable at $\bfone_{\tI}$, (\ref{k approximation}) does not hold. Non-polyhedral definitions of differentiabiilty are relevant when $\tI$ is not finite.
\end{remark}

\subsection{Euler Gaps and the Core}

\begin{definition} The core of $G:\bfone_{\mathcal{S}} \rightarrow \mathbb{R}_+$ consists of $r\in \mathbb{R}^n_+$  such that 
\[  \partial_C G(\bfone_{\tI}) = \{r : r\cdot \bfone_{\tI} - G(\bfone_{\tI}) \geq r\cdot \bfone_S - G(\bfone_{S}), \quad \forall S \subset \tI\} \] 
\end{definition}

Such an $r$ exists if and only if $G$ is self-concavifying at $\bfone_{\tI}$:
\[ G(\bfone_{\tI}) = F(\bfone_{\tI}) = \sup\, \{\sum_k \alpha_{S_k} G(\bfone_{S_k}) : \sum_k \alpha_{S_k} \bfone_{S_k} = \bfone_{\tI}\} \]
called a {\em balanced game} (\cite{Shapley1967})
The game $G$ is self-concavifying or {\em totally balanced\/} if for every $S$ there exists an $r$ such that
\[  r\cdot \bfone_S - G(\bfone_{\tI}) \geq r\cdot \bfone_T - G(\bfone_{T}), \quad \forall T \subset S \]
I.e., $G= F$ on $\bfone_{\mathcal{S}}$.

Therefore, the characteristic function  of a totally balanced game implies that  $G = F:\bfone_{\mathcal{S}} \rightarrow \mathbb{R}_+$,  a restricted version of $F$.   
Evidently,
\[ \partial F(\bfone_{\tI}) \subseteq \partial_C F(\bfone_{\tI})\]
Note: $\partial_{\text{C}} F(\bfone_{\tI}) =\{r\}$ implies $\partial F(\bfone_{\tI}) = \partial_C F(\bfone_{\tI})$.

\begin{definition}  The {\em equal treatment core\/} for $k\bfone_{\tI}$ as
\be  \partial_{\text{eC}} F(k\bfone_{\tI}) :=\{r :  r\cdot k\bfone_{\tI} - F(k\bfone_{\tI}) =0 \geq r\cdot x - F(x), x = \sum_k \alpha_{S_k} \bfone_{S_k} \leq k \bfone_{\bfone_{\tI}}, \alpha_{S_k} \in \{1,2, \ldots \}\} \ee
%(Evidently, $\partial_{\text{eC}} F(k\bfone_{\tI}) = \partial_{\text{C}} F(k\bfone_{\tI}) $ when $k=1$.)
\end{definition}

Since the equality $r\cdot k\bfone_{\tI} - F(k\bfone_{\tI}) =0$ does not vary with $k$,  the effective  number of inequalities increases with the number of allowable $x$ in (39). And the greater the number of inequalities, the more restricted is  $\partial_{\text{eC}} F(k\bfone_{\tI})$. Therefore, 
\be  \partial F(\bfone_{\tI}) \subseteq  \partial_{\text{eC}}\, F((k+1)\bfone_{\tI}) \subseteq \partial_{\text{eC}}\, F(k\bfone_{\tI})\ee
The larger is $k$ the closer the inequalities in  $\partial_{\text{eC}}\, F(k\bfone_{\tI})$ approximate the inequalities in  $\partial F(\bfone_{\tI})$.

\begin{definition} There is {\em core equivalence\/} for $\partial F_{\text{eC}}\, F(k\bfone_{\tI})$ when $\partial F_{\text{eC}}\, F(k\bfone_{\tI})$ satisfies the weak version of Euler's Theorem, i.e., when 
$\partial F_{\text{eC}}\, F(k\bfone_{\tI}) =  \partial F(\bfone_{\tI})$. 
\end{definition} 

\begin{remark} {}  {\em Core equivalence\/} is normally associated with an equality between the core of a cooperative game based on an economic model of exchange and the utilities individuals receive from the Walrasian equilibrium allocations of that model. The above definition is consistent with that terminology.
\end{remark}

\subsection{Euler Gaps and the Shapley Value}

$\partial F(\bfone_{\tI})$ exhibits some, but not all, of the conditions defining the Shapley value. 

$\partial F(\bfone_{\tI})$ is efficient: $r\in \partial  F(\bfone_{\tI})$ implies $r\cdot \bfone_{\tI} = r'\cdot \bfone_{\tI} = F(\bfone_{\tI})$. 

$\partial F(\bfone_{\tI})$ is additive: $\partial (F_1 + F_2)(\bfone_{\tI}) = \partial F_1 (\bfone_{\tI}) +  \partial F_2 (\bfone_{\tI})$. 

However, multiplicity of valuations in $\partial F(\bfone_{\tI})$ allows for the failure of the anonymity and carrier conditions for any arbitrarily selected $r \in \partial F(\bfone_{\tI})$.  

Without multiplicity,

\begin{proposition} 
If $\partial F(\bfone_{\tI}) = \nabla F(\bfone_{\tI})$, the Shapley value of $\Phi(S)\sim F(\bfone_S)$ coincides with  Euler's Theorem. 
\end{proposition}

If $\partial F(\bfone_{\tI})\neq \nabla F(\bfone_{\tI})$, the Shapley formula has to be applied to the weak version of Euler's Theorem to obtain  $\nabla F^{\text{Sh}}(\bfone_{\tI})$.

\subsection{Two-person Games}

When $n=2$, $\Psi(u) \sim F\sim\psi (U) \sim G$ is self-concavifying. Two-person games exhibit the properties of a positively homogeneous saddle-function. Restating Proposition 2 in the terminology of this Section: 
The saddle-function exhibits Euler's Theorem if and only if the game is constant-sum. If the 2-person game is not constant sum, the Euler Gap  $E_k(\bfone_{\tI})$ is a positive constant equal to gains from cooperation for all $k =1,2,\ldots$. If $n\geq 3$, $G \neq F$.

A constant sum game is defined by a $U_c$ such that
\be \bfone_{\tI} U_c \cdot \mfz = c, \forall \mfz \in \mfZ_{\tI}\ee
implying that all $\mfz$ are efficient. Denote the set of $n$-person  games with utility sums equal to $c>0$ as
\[ \mathbf{U}_c = \{U_c: \bfone_{\tI} U_c \cdot \mfz = c, \forall \mfz \in \mfZ_{\tI}\}\]
When $n=2$, 
\[\bfone_{\tI} U_{2c}(a_1,a_2) = u_1(a_1,a_2) + u_2(a_1,a_2) = u_1(a_1,a_2) +[c- u_1(a_1,a_2)], \forall (a_1,a_2) \in A_1\times A_2\]

The  saddle-point for $(\bfone_{\tI},U_{2c}$) is $\mfz(\bfone_{\tI}, U_{2c})$ such that
\be     [\mfz(\bfone_{\tI},U_{2c})]^t[U']^t\cdot \bfone_{\tI} \geq  [\mfz(\bfone_{\tI},U_{2c})]^t U^t\cdot \bfone_{\tI} =   \bfone_{\tI}U_{2c}\cdot \mfz(\bfone_{\tI},U_{2c})\geq \bfone_{\tI}U_{2c}\cdot \mfz,  \forall U'_{2c}\in \mathbf{U_{2c}}, \forall \mfz\in \mfZ_{\tI}, \ee
an equivalent version of the minimax condition.

\section{Euler Gaps for $H_x$ and Incentives} 

The transferable utility assumption in cooperative games has been a starting point for the study of incentives, the focus of attention in mechanism design.  Incentives to misrepresent  utilities are related to the properties of $H_x(U)$,  holding $x$ fixed.

The directional derivative of $H_x$ at $U$ in the direction $D$ is 
\be H_x(U)(D) = \lim_{\lambda \searrow 0} \frac{H_x(U +\lambda D) - H_x(U)}{\lambda} \ee

In parallel with the directional derivative of $F_U(x)(d)$ whose properties mimic $F_U$, 
\begin{fact}  $H_x(U)(D)$ is positively homogeneous and subadditive, therefore convex, in $D$.
\end{fact}

Also in parallel with $F_U(x)(d) = \inf\, \{r\cdot d: r\in \partial F_U(x) \}$,
\begin{fact}
\[ H_x(U)(D) = \sup\, \{xD\cdot \mfz: \mfz \in \partial H_x(U)\}\]
\end{fact}

\subsection{Incentive compatibility for Discrete and Infinitesimal Individuals} 

Denote by $U(u'_i)$ the direction $D$ in which the $i^{th}$ row of $U$ is replaced by $u'_i$.  At $x = \bfone_{\tI}$,
\[ H_{\bfone_{\tI}}(U)(U(u'_i)) = \sup\, \{\bfone_{\tI} U(u'_i)\cdot \mfz: \mfz \in \partial H_{\bfone_{\tI}}(U)\}\]
Since for incentives,  $\bfone_{\tI}$ remains fixed, write $H_{\bfone_{\tI}}(U)(\cdot)$ as $H(U)(\cdot)$ and $\mfz(\bfone_{\tI},U)$ as $\mfz(U)$

If $u_i$ were to report $u'_i$, its maximum possible gain would be 
\[ \bar{r}_i(U(u'_i); u_i) : = \sup_{u_i'}\, \{ u_i\cdot  \mfz(U(u'_i)) - m_i( \mfz(U(u'_i)): \mfz(U(u'_i)) \in \partial H(U(u_i'))\}, \]
an outcome $u_i'$ could receive, as evaluated by $u_i$. 

From total utility maximizing choices, 
\[\mfz(U)  \in \partial H(U) \Longrightarrow U\cdot \mfz(U) \geq U\cdot \mfz(U(u'_i)), \forall u_i'\]
Hence, if $i$ were to gain from misrepresentation by receiving a more favorable distribution of the gains, it would be at the cost of reducing overall gains. 

At $U(\cdot)$, the most favorable outcome for $u_i$ is
\[ \bar{r}_i(U(\cdot); u_i) = \sup_{u'_i}\, \bar{r}_i(U(u'_i); u_i)\]

\begin{definition} $\mfz(U)$ is incentive compatible for $i$ at $U$ if and only if
\[ r_i(U) = u_i\cdot \mfz(U) - m_i(z(U) = \sup_{u'_i}\, \bar{r}_i(U(u'_i); u_i)\]
\end{definition} 

\begin{proposition} 
If $i$ receives all the gains it contributes, it cannot gain by misrepresenting
\[ r_i(U) =  -\Delta_k F(\bfone_{\tI})(-\bfone_i) = 
u_i\cdot \mfz(\bfone_{\tI},U) - m_i(\mfz(\bfone_{\tI},U)) \geq  u_i\cdot \mfz(\bfone_{\tI},U(v_i)) - m_i(\mfz(\bfone_{\tI}, U(v_i)))\]
\end{proposition} 

\begin{corollary} If the Euler Gap  at $U$ is $E_k(\bfone_{\tI}) =0$, no individual can gain from misrepresentation.
\end{corollary}

In other words, the Euler distribution passes a non-cooperative test of incentive compatibility for an efficient distribution.  

Conversely, if the Euler gap is positive, an individual can gain from misrepresentation: \[
\begin{aligned}
r_i(\bfone_{\tI}, U) & <  -\Delta_k F(\bfone_{\tI})(-\bfone_i)\\
&  \Longrightarrow \exists v_i \quad 
 u_i\cdot \mfz(\bfone_{\tI},U(v_i)) - m_i(\mfz(\bfone_{\tI}, U(v_i))) > u_i\cdot \mfz(\bfone_{\tI},U) - m_i(\mfz(\bfone_{\tI},U))
\end{aligned} 
 \]

$(k\bfone_{\tI},U)$ is incentive compatible if (and only if) $E_k = 0$.

 At the discrete margin, $\partial F_U(\bfone_{\tI})$ may exhibit the property that 
for any $r\in \partial F_U(\bfone_{\tI})$,  
\[  r_i < -\Delta F_U(\bfone_{\tI})(-\bfone_i) \forall i\] 
I.e., all individuals can gain by misrepresentation. 

At the infinitesimal margin, for any $i$ there exists $r \in  \partial F_U(\bfone_{\tI})$ such that
\[ \bar{r}_i = -F_U(\bfone_{\tI})(-\bfone_i) \]
In that case, if $i$ receives $\bar{r}_i = u_i\mfz(U) - m_i(z(U))$,  $i$ would not gain from misrepresentation.  

\begin{corollary} If $\nabla F_U(\bfone_{\tI}) \neq \emptyset \Longleftrightarrow \nabla H_{\bfone_{\tI}}(U)\neq \emptyset$, no infinitesimal individual can gain from misrepresentation.
\end{corollary} 

The example of $n=2$ when $G=F$ and $[-\Delta F(\bfone_{\tI}(-\bfone_1)] +  [-\Delta F(\bfone_{\tI}(-\bfone_2)] > F(\bfone_{\ti})$ implies that $E_{\infty} > 0$. For all such games, $\partial F(\bfone_{tI})$ and $\partial H(U)$ are multi-valued. If  $i$ is not receiving its maximum gains, a small change, $|u_i - u_i'| < \epsilon$,  from $U$ to $U(u_i')$ can improve the gains to $u_i$.% ...Glove game...Edgeworth's master-servant example....neither of these examples are used to exhibit manipulation ?? 

\subsection{non-TU Euler Gaps}

Let $\Gamma = \{\gamma = \nvec{\gamma_i}: \gamma_i \geq 0, \sum_i \gamma_i =1\}$. 

From the positive homogeneity of $H_x(U)$, $H_x(\gamma U), \gamma \in \Gamma$ is a normalization of the set all weighted variations of $U$. For $x = \bfone_{\tI}$, $U$, and $\gamma \in \Gamma$,
\[ \mfz(\gamma) \in \partial H_{\bfone_{I}} (\gamma U) = \partial H_{\bfone_{tI}} (\gamma U), \quad r(\gamma) \in  \partial F_{U} (\bfone_{\tI}) = F_{\gamma U} (\bfone_{\tI}) \] 

The money payments connecting $\mfz(\gamma)$ and $r(\gamma)$ are
$m_i(\gamma) = u_i\cdot \mfz(\gamma) - r_i(\gamma)$, which was first noted in the game context by Shapley (\cite{shapley1969utility}). Moreover, he applied the following fixed-point argument to ensure the money payment to be zero.

Mapping $M: \Gamma \rightrightarrows \Gamma$
\[ M(\gamma) = \frac{ \gamma_i + m_i(\gamma)^+}{1 + \sum_i m_i(\gamma)^+} \]
is a convex, closed-valued correspondence. Its fixed-points imply that $m_i(\gamma) = 0, \forall i$. I.e, there exist utility weights for $U$ such that $\gamma_i  u_i\cdot \mfz(\gamma) = r_i(\gamma), \forall i$.

There may be multiple fixed points with total utilities depending on which $\gamma$ are chosen as a fixed point.

The non-TU fixed point satisfies Euler's Theorem if the correspondence is single-valued at the fixed point:
\[\mfz(\gamma) = \nabla H(\gamma) \Longleftrightarrow r(\gamma) = \nabla F_{\gamma}(\bfone_{\tI}) \]

\begin{remark} \sc{(An Added Level of Duality)}  A formulation using decentralized (Lindahl) price-taking, 
in which money payments are defined by personalized prices for public goods, can be shown to yield equivalent conclusions with respect to incentive compatibility, both for TU and non-TU models. This, along with comparable results for private goods models, underlies  Vickrey-Clarke-Groves conclusions for public and private goods  in mechanism design.  
\end{remark}

\section{Positively Homogeneous Saddle-Functions for Exchange}

The purpose of this section is to demonstrate an essentially one-to-one correspondence between  positively homogeneous saddle-functions in games and their counterparts for (polyhedral) models of exchange economies. As is well-known, concavity of individual utility functions for private goods lends itself to ``self-concavity'' without invoking the qualification in Section 4.2 needed for that property in games. 

To emphasize similarities the same $A = A_1\times A_2 \cdots \times A_n$ in games will be the basis for exchange. $E_i(a_i) = \nvec{E_i(c,a_i)} \in \mathbb{R}^{\ell}$ is a commodity vector of positive and negative excess demands for commodities $c = 1,\ldots, \ell$ constituting the trade by $i$ using activity $a_i$. The matrix of trades for $i$ based on available activities is $E_i \in \mathbb{R}^{\ell} \times R^{A_i}$.

 The set of convex combinations of $A_i$ is $Z_i = \{z_i: z_i(a_i) \geq 0, \sum_{a_i} z_i(a_i) =1\}$. 
 \[ E_i[Z_i] = \{E_i z_i = \sum_{a_i} E_i(a_i)z_i(a_i) \in \mathbb{R}^{\ell}, z_i \in Z_i\}\]
is the convex set of possible trades available to $i$, with $\bfzero \in  E_i[Z_i]$ indicating that no-trade is feasible. 

The set of possible trading activities  $E= \nvec{E_i}$, the $A_i$ trading activities for each $i$, is fixed throughout. 
The set of feasible trades is determined by the population  $x$ as
\be Z(x) = \{ z = \nvec{z_i} \in Z: xEz = \sum_i  x_i[\sum_{a_i} E_i(a_i) z_i(a_i)] = \bfzero\} \ee
as those trades that for which the sum of population (excess) demands with (excess) supplies are equated.

A utlity vector $v_i \in \mathbb{R}^{\ell}_+$ evaluates trades for $i$ as
\[ v_i\cdot E_i z_i  = v_i E_i \cdot z_i = \sum_{a_i} \sum_c v_i(c) E_i(c,a_i) z_i(a_i)\]
The utilities resulting from each $i$ choosing $a_i$ is
\[ VE(a) = \nvec{v_iE_i(a_i)} \in  \mathbb{R}^n\]

For $z \in Z = \times Z_i$, utilities are
\[ VE\cdot z = \nvec{v_iE_iz_i} \in   \mathbb{R}^n\]

The parameters for exchange are $(x,V)$, the counterparts to $(x,U)$ for games. In both cases, utilities  may vary, but not the set of opportunities themselves. 

The counterpart of $xU\cdot \mfz = \mfz^tU^t \cdot x$ is
\[ xVE\cdot z = z^tE^tV^t \cdot x \in \mathbb{R} \qquad  xU\cdot \mfz = \mfz^tU^t \cdot x\]

The counterpart of $F_U(x) = \sup_{\mfz}\, \{ xU\cdot \mfz - \delta_{\mfZ_x}(\mfz)\}$ is
\be \mathbf{F}_V(x) = \sup_z\, \{xVE\cdot z - \delta_{Z(x)}(z)\} \ee

The counterpart of $H_x(U) := \inf_{\mfz}\, \{ xU\cdot \mfz - \delta_{\mfZ_x}(\mfz)\}$ is 
\be \mathbf{H}_x(V) = \inf_z\, \{xVE\cdot z - \delta_{Z(x)}(z)\} = \inf_z\,  \{ z^tE^tV^t\cdot x - \delta_{Z(x)}(z)\} \ee

And
\[z(x,V) \in \partial \mathbf{H}_x(V), \qquad r(x,V) =  z(x,V)^tE^tV^t \in \partial \mathbf{F}_V(x)\]

It is asserted without proof that each of the conclusions for the saddle-function for games based on $F_U(x)$ and $H_x(U)$
applies to the saddle-function based on $\mathbf{F}_V(x)$ and $\mathbf{H}_x(V)$ for exchange.

\section{Concluding Remarks}

Most of the above results calling attention to differentiability and positive homogeneity are special cases of  conclusions previously established in more general settings. Nevertheless, the elementary model examined here is regarded as exhibiting salient features of more general models. 

The role of large numbers has been defined for a continuum of individuals consisting of an infinite variety of types.  Although often included in assumptions underlying the model, a continuum of individuals does not necessarily imply the Euler condition that infinitesimal individuals necessarily have no influence. For example, in the TU case, multiplicity in the set of core solutions that is shown to coincide with a corresponding multiplicity of  Walrasian allocations implies that infinitesimal individuals can influence the outcome. A similar conclusion applies for the non-TU case. However, if the preferences/utilities of individuals are differentiable, the Euler condition applies. 

Relatedly, research on models with a continuum of individuals emphasizes the existence of large numbers of infinitesimal individuals, {\em per se}, without drawing equal attention\,---\,although it is certainly recognized, to the positive homogeneity property of such models. It is as if it were an unavoidable conclusion of large numbers that per capita gains can be maximized when  self-sufficient trading groups can be confined to a scale that is arbitrarily small compared to the overall population.  Such a counterfactual to the principle that ``specialization and division of labor is limited by the extent  of  the market'' would not apply if the heterogeneity of types increased with the number of individuals, precluding constant returns/positive homogeneity with respect to individuals.

\end{document}